\documentclass[a4paper,11pt]{article}
\usepackage{jheppub} 
\usepackage{lineno}
\usepackage{tikz}
\usepackage{slashed}

\arxivnumber{2608.08972} 

\title{\boldmath Matrix Multiverses Meet Multiple Mythologies}

\author[a]{Sidan A\note{Corresponding author.}}
\author[a]{, Tom Banks}
\author[b]{, Willy Fischler}
\affiliation[a]{Department of Physics and NHETC,
Rutgers University, \\
Piscataway, NJ 08854 USA}
\affiliation[b]{Department of Physics and Astronomy and Weinberg Institute, University of Texas, \\ Austin, TX 78712 USA}

\emailAdd{sidan.aa@rutgers.edu, tibanks@ucsc.edu, fischler@austin.utexas.edu }

\abstract{We present a model in which asymptotically de Sitter universes of various types, and de Sitter (dS) radii $R_n$, live in the interiors of black holes in a maximally entropic flat $p = \rho$ Friedmann-Robertson-Walker universe.  These dS universes clearly have many quantum states.  We argue that they decay and equilibrate with the maximal entropy universe on time scales of order $\alpha_n R_n {\rm ln} (R_n /\delta_n)$, where $\alpha_n$ are model dependent dimensionless constants and $\delta_n$ are the initial distances between the shell, satisfying the Israel junction conditions, and the dS horizon. These are time-scales as viewed by a detector following a trajectory far from the would be cosmological horizon. These times are all exponentially shorter than dS recurrence times, which have no meaning in this model.  It is impossible for a detector inside one of the dS universes to determine whether it is actually part of such a structure.  This model could be used as the basis for claiming that certain constants of nature or cosmological initial conditions were chosen to have mathematically unnatural values because other values could not lead to any form of intelligent life. }

\begin{document}
\maketitle
\flushbottom

\section{Introduction: A Model of Everything and More}

Our basic model will consist of $N^* (N^* + 1)$ $1 + 1$ dimensional Dirac fermion fields, living on an interval $[0,\pi]$ with $15 - 20$ spatial momentum modes. The fields may be thought of as transforming in the first $N^*$ spinor spherical harmonics on the two sphere.  $N^*$ is an EXTREMELY LARGE INTEGER,  which we can imagine taking to infinity at the end of our considerations.  To fix ideas about how large it is: it should contain many multiples of $10^{123}$.  We will in fact need an infinite number of copies of this system, but they will be used to form a {\it non-isometric encoding} \cite{noniso} of quantum information into the Hilbert space of a single copy.  The purpose of this encoding is to make causal time evolution, unitarity, and the covariant entropy bound \cite{fsb,fsb1,fsb2,fsb3} consistent with each other.  

In the work of \cite{tbwflm} we showed that a model of this type could be used as the basis for a finite construction of a Big Bang cosmology with scale factor $a(t) \sim \sinh^{1/3} (3t/(N^* L_P))$, where $L_P$ is the Planck length.   In each causal diamond of some geodesic starting at the Big Bang and going up to cosmic time $\sim t$, we fix the density matrix inside the diamond by choosing the modular Hamiltonian to be 
\begin{equation} 
    K(t) = L_0 (t) + \frac{g^2}{t} \int {\rm Tr} J^m J_m , 
\end{equation} 
where $L_0 (t)$ is the Virasoro generator of $t (t + 1)$ of the Dirac fields, $g$ is the coupling constant, and $J^m$ is the $U(t)$ current constructed from those fields.  We define the energy in the diamond, as measured along the geodesic, as $E = \langle K(t)\rangle /t$, which takes into account the redshift of states living at the maximal area surface on the boundary of the diamond.  Since the volume for flat FRW cosmology scales like $t^3$ and $<K(t)>$ scales like $t^2$, this gives the proper Friedmann scaling of energy, and the relation between entropy density $\sigma$ and energy density $\rho$ is 
\begin{equation} 
    \sigma \sim \sqrt{\rho} , 
\end{equation} 
characteristic of the maximally stiff $p = \rho$ fluid, as long as $t < N^* L_P $. After that time the number of fields does not change.  A time evolution that guarantees this evolution of density matrices is the analog of {\it half sided modular inclusion} from Algebraic Quantum Field Theory (AQFT)\footnote{Half Sided Modular Inclusion is a one parameter group of automorphisms generated by the difference between the modular Hamiltonians of two closely spaced nested diamonds. We instead use a semi-group in which each time step is generated by the difference between the modular Hamiltonians of two diamonds in a nested covering of the maximal diamond along a complete geodesic. For an asymptotically dS space the maximal diamond Hilbert space has finite dimension and there are no delicate limiting questions involved.}, using these finite density matrices instead of the QFT modular operators.  

That is, at proper time $t_n$ since the singularity (the time when there's only one fermion field in the causal diamond and the diamond area is small in Planck units) we take the time dependent Hamiltonian to be
$H (t_n) = t_n^{-1} [K (t_n) - K(t_{n-1}) + g(t_n)] $ where $g(t_n)$ is chosen randomly at each time from the Lie algebra of the large symmetry group of $K_n$.  

If we continue the time evolution once $t = N^*L_P$ using just $g(t_n)$ then we get a model that realizes the conjecture \cite{tbwfds,tbwfds1,tbwfds2} that de Sitter (dS) space has a finite dimensional quantum Hilbert space.  The original conjecture is modified by replacing the maximally uncertain density matrix on that space with the generalized Carlip-Solodukhin ansatz.  Recent evidence in favor of that replacement can be found in \cite{tbpdds}.  

The natural time scales in this evolution are designed to be those in causal diamonds.  Later, when we explore constrained states of models like this in which the initial state of a large matrix of fields is forced to be such that square bilinear $N \times N$ matrices are block diagonal, with of order $N$ vanishing elements, the natural time scale for thermalization of these vanishing elements is, by construction, of order $N$ in Planck units. The all to all nature of the interactions generated by $g(t)$ imply that this system is a fast scrambler, with a scrambling time of order $N {\rm ln} N$ \cite{Susskind:2005js, Hayden:2007cs, Sekino:2008he}. Intuitively this is because, thinking of the fermions as spinors on the two sphere, the bilinears are differential forms, and the unitary transformations on the fermions are a fuzzy approximation to the group of area preserving maps.  

The reason for insisting on an infinite number of copies of the model, one for each of a discrete infinity of geodesics in the background $p = \rho$ space-time,  is to impose unitarity on our quantum gravitational system. For finite dimensional Hilbert spaces, half sided modular inclusion is a unitary embedding of a smaller Hilbert space into a larger one.  In order to boost it into a full unitary transformation we use the idea of {\it non-isometric embedding} \cite{noniso}. We do quantum mechanics separately on each geodesic of the background space time, with the connection on this Hilbert bundle given by the Quantum Principle of Relativity, which states that the Hilbert space of an overlap diamond between any two diamonds on different geodesics is a tensor factor in each of the individual diamond Hilbert spaces.  The entanglement spectra of the density matrices on these two tensor factors must be the same, as computed by either geodesic's quantum mechanics.  In our model, this is guaranteed by using identical quantum systems on each geodesic, and by the $U(t) \times U (t + 1)$ symmetry of the density matrices.  We can think of this as an approximation to the group of area preserving maps of the two sphere.  Our model guarantees that every diamond has a density matrix completely specified by its total area.  

Given this basic model, we can now build a {\it multiverse} by starting with initial conditions where the bilinears in fermions are all block diagonal matrices, with blocks of different sizes.
The quantum mechanics on different geodesics will start on different blocks and incorporate different blocks at different times in their evolution.  Let us first assume that all but one of those blocks is very large in Planck units, but $\ll N^*$ with the remaining block of order $N^*$ in size.  The large block evolves as a $p = \rho$ universe on a time scale where the other blocks have reached their dS asymptotes with radii $R_n$.  Now suppose the expanding horizon of the large block incorporates one of those smaller blocks.  The small block is a large entropy equilibrium system, with modular fluctuations satisfying the Carlip-Solodukhin law.  It is smaller than the particle horizon size in the background universe. On the one hand we've demonstrated that it behaves like an asymptotic de Sitter (AsdS) universe. On the other hand it's some kind of high entropy localized excitation inside the $p = \rho$ universe.
The geometrical-hydrodynamic approximation to it is obviously a dS horizon glued in to the interior of an approximately Schwarzschild black hole in the $p = \rho$ universe.
We will describe the appropriate Israel junction conditions below.

The evolution of the system in the matrix model is clear.  The time dependent unitary transformation mentioned above will excite the off diagonal degrees of freedom and bring the system to the Carlip-Solodukhin equilibrium state for the large block that includes the original block diagonal system.  This is dual, in the GR picture, to the following processes.  The Israel junction condition glues the AsdS universe across a time-like boundary near its dS horizon into the interior of a larger black hole in the $p = \rho$ geometry. If we allow energy flow into the black hole, the time-like boundary collapses to a singularity.  We interpret this collapse as thermalization of the degrees of freedom of the AsdS universe with those of the black hole, following \cite{tbsing}.  That paper was based on many works by two of the present authors, which modeled in-fall of matter to a black hole in terms of excitation of off diagonal blocks of a block diagonal matrix model state, and on the fact that in-fall times and thermalization times coincide parametrically for all known black holes.   

In a companion paper \cite{galaxyseeds} to this one we have shown how similar considerations inside one of the smaller blocks, could lead to a universe that underwent periods dominated by ordinary radiation and matter, fitting the spectrum of cosmic microwave background (CMB) fluctuations and perhaps even the distribution of Primordial Black Holes necessary to understand the early history of galaxy formation.  Viewed from the point of view of the present model, the initial conditions for that development were rather improbable.  However, the huge value of $N^*$ we are postulating gives many copies of asymptotically dS universes with cosmological constants (c.c.) similar to the one our universe appears to have, so the environmental constraints necessary for the evolution of organized life, whatever those may be, might play a role in understanding why the world looks the way it does.  

Apart from such speculations, one clear virtue of our model is that it gives a mathematically precise definition of closed, asymptotically de Sitter universes, which does not obey many of the speculative conjectures that have been made in the literature about such systems.  Our model may not be correct, but it is mathematically well defined and has features that mirror the semi-classical physics of models of dS space. This is in contrast to field theory models of inflation which are manifestly ill defined as they involve length scales much smaller than the Planck scale without having a theory of quantum gravity.  The companion paper shows that our model also exhibits features that resemble the universe we inhabit. 

\section{The Number of States of Closed Universes, Boltzmann Brains, and Other Philosophical Matters}

Despite some recent claims, evidence seems to point to an asymptotically de Sitter future for the universe we inhabit.  Since the seminal work of Gibbons and Hawking \cite{GH}, most physicists have concluded that such a universe has a finite entropy.  Two of the current authors \cite{tbwfds,tbwfds1,tbwfds2} postulated that this implied a finite dimensional Hilbert space with the Gibbons-Hawking entropy equal to the logarithm of the dimension.  A more correct conjecture for the density matrix follows from the work of Carlip \cite{carlip} and Solodukhin \cite{solo}.  The generalization of their ansatz to dS space appeared in \cite{BZ} and is supported by \cite{tbpdds}.  A sort of derivation of it can be obtained \cite{tbwflm} by viewing dS space as the asymptote of a flat FRW cosmology, which saturates the covariant entropy bound \cite{fsb,fsb1,fsb2,fsb3} at all times.  

The static patch of dS space has a time-like Killing vector, which many physicists (including two of the authors at times) have taken to describe a time independent Hamiltonian evolution operator in the static patch of the universe.  This raises the issue of recurrences, which in their most picturesque form \cite{BB} are called Boltzmann brains.  This is an ancient problem, first raised by Boltzmann's assistant Schultz.  

The first point that one should make clear is that there is no way, even in principle, for any sort of experiment to observe these recurrences.  This is a consequence \cite{tbwfsp} of the combination of simple properties of quantum measurement theory, quantum field theory, and general relativity.  Quantum measurement theory says that in order to measure a q-bit of a microscopic system, one must entangle it with a ``semi-classical q-bit" of a measuring device.  A semi-classical q-bit is a two valued collective variable of a large quantum system possessing $\mathcal{O}(N)$ microscopic constituents, with $N > 10^{20}$, having the following two properties
\begin{itemize}
\item The quantum fluctuations of the semi-classical q-bit are $\mathcal{O}(N^{-p})$ with $p$ of order one.
\item The interference terms in histories of the semi-classical q-bit's evolution are $\mathcal{O}\left(e^{- c N}\right)$ with $c$ of order one.  
\end{itemize}  These are the conditions for observations of the semi-classical q-bit to give robust results, which can be interpreted via the classical theory of probability.

The only systems in the real world with these properties are semi-classical q-bits described mathematically by (cut off) quantum field theory (QFT).  However, it has been known since at least the late 1980s \cite{tHooft} that in a causal diamond of area $A$, at most $\sim (A/L_P^2)^{3/4}$ of the states can be described by quantum field theory\footnote{Cohen, Kaplan and Nelson \cite{cknetal,cknetal1,cknetal2,cknetal3} showed that leaving those states out of QFT did not interfere with its agreement with experiment.}.  The maximal entropy states localized near the geodesic are all black holes, which have very few semi-classical q-bits because of fast scrambling \cite{Sekino:2008he} of quantum information.  Even worse, the Schwarzschild-de Sitter black hole entropy formula tells us that the more massive and robust you make the detector, the fewer states of the dS universe can actually be accessed.  Finally, any detector with a Schwarzschild radius much smaller than $R_{dS}$ can only stay on a static geodesic for a time that is a dimension dependent power law in $(R_{dS}/L_P)$ \footnote{This assumes the center of mass of the detector is a robust semi-classical variable, which gets kicks from Gibbons-Hawking gravitons, causing it to perform a random walk from the origin to the horizon of any given static patch.}, so the static time evolution does not correspond to the clock on any real detector.  The most robust objects in a dS universe are local clusters of galaxies.  Each of these has its own idiosyncratic non-geodesic trajectory.  It turns out that the lifetime of local clusters is bounded by roughly the same time scale, because of simple gravitational physics.  The in-fall time governing the timescale for collapse to a black hole is governed by the local over-density  and its size.  However, if the size is too large, the Hubble flow beats the local gravitational attraction and the cluster is not bound.  In our universe, the collapse time for the largest known superclusters is somewhere between $10^{20} - 10^{30}$ years.  This is much longer than the dS Hubble scale, but extremely short compared to estimates of recurrence times.  

In fact, there is no real argument that the dS Killing vector is a time independent Hamiltonian in quantum mechanics.  Diffeomorphisms are gauge transformations and there is no asymptotic time-like or null boundary on which to define asymptotic global symmetries.  We have just explained why attempts to base a time independent Hamiltonian formalism on a detector in the space-time does not make sense for finite $R_n/L_P$.  If we examine the models \cite{tbwflm} for maximally entropic asymptotically dS space-times, we can in fact see that there is no reason for the asymptotic Hamiltonian to be time independent.  Those models originated as models that maximized entropy on the Hilbert space of a finite number of fermion oscillators by having a modular Hamiltonian for a causal diamond of horizon size $\sim t$, which was a small perturbation of a random quadratic Hamiltonian in $\mathcal{O}(t^2)$ fermions.  For large numbers of fermions this becomes the Hamiltonian of Dirac fermions, which we arranged as a $t \times (t + 1)$ matrix of cutoff $1 + 1$ dimensional Dirac fields, with a $U(t) \times U(t+1)$ invariant current-current perturbation.  Nothing prevents us from changing the modular Hamiltonian by a time dependent symmetry transformation as we evolve past the point where the matrix size no longer increases with $t$ (corresponding to the dS asymptote in space-time).  Such a system will still have the hydrodynamic properties of dS space, finite entropy and fast scrambling, and approximate Carlip-Solodukhin fluctuations of the modular Hamiltonian, but no recurrences, because many properties of the system are still fluctuating randomly in a time dependent manner.  

A second fashionable claim about ``closed universes like dS space"  is that they have a one dimensional Hilbert space \cite{1dbabyu,1dbabyu1,1dbabyu2,1dbabyu3}.  There are many reasons to be suspicious of this claim, one of which is that it would seem to preclude any attempt to view the well established quantum theories of gravity for vanishing cosmological constant as limits of the positive c.c. models.  In the present paper, we will show that both these claims, whatever their theoretical validity, cannot be argued to have anything to do with the universe we live in, on the basis of the observation that it appears to be approaching dS space asymptotically.  We will show that a horizon volume of any asymptotically dS space can be embedded inside a McVittie black hole in the flat $p = \rho$ FRW space-time.  The $p =\rho$ spacetime has causal diamonds of arbitrary size, and so its conformal boundary is null and has maximal area that is infinite.  The black hole has a finite area event horizon which must be larger than or equal to that of the dS space.   It is an object with finite entropy.  Thermodynamics then implies that the finite entropy object is unstable to having its degrees of freedom equilibrate with those of the larger system.  

We will argue that the lifetime of the asymptotic dS universe depends on conditions that cannot be measured by observations inside that universe before the decay begins to occur, but that it is always much shorter than recurrence times in the dS universe.  This model demonstrates both a context in which a ``closed" dS universe has a large number of quantum states, and one in which dS recurrences never occur. It is based on the Israel junction condition in General Relativity, so cannot be dismissed as a consequence of unfounded speculations about the nature of quantum gravity.  

In addition to this intrinsic instability, if we take two such embedded dS spaces, perhaps with different dS universes, then depending on their initial conditions in the embedding space, they may collide and merge, completely changing the configuration in the interior.  This might happen at any proper time along trajectories in the interior, so might happen before any recurrences.  And there is no way to know, from measurements in the interior, whether and when this catastrophe will occur.  

The other virtue of this model, our original motive for constructing it  \cite{tbwfhstinf}, is that it provides a mechanism for environmental selection of the c.c..  That is, it is, within the realm of finite quantum mechanical models that obey the covariant entropy bound (CEB), a multiverse, which contains an arbitrarily large number of asymptotically dS universes, with lifetimes of order the Hubble radius or longer and arbitrary values of the c.c. and different cosmological initial conditions.  So it provides a framework for alleging that our universe might have unusual properties because they were necessary for the evolution of life that could study it.  Like all such models, it suffers from the problem that its details cannot be verified by experiments done within a single universe, at least until a catastrophe that probably brings about the extinction of life.

\section{A Holographic Multiverse, Satisfying the Covariant Entropy Bound}

The flat Friedmann-Robertson-Walker cosmology with scale factor
\begin{equation} a_n (t) = \sinh^{1/3} (3t/R_n) , \end{equation} saturates the covariant entropy bound \cite{fsb,fsb1,fsb2,fsb3} for all times and all positive $R_n$ .   For finite $R_n$ it has a cosmological event horizon, with Gibbons-Hawking entropy $\pi (R_n / L_P)^2$.  The infinite $R_n$ limit is the FRW cosmology with maximally stiff $p =\rho$ equation of state.  

We can match a horizon volume of any these cosmologies into the $R = \infty$ limiting cosmology by using the Israel junction condition \cite{Israel}, to paste it into the interior of a McVittie black hole of Schwarzschild radius $R_s$ along a time-like boundary $\partial B$. The stress tensor on the boundary satisfies the dominant energy condition as long as $R_s \geq R$. Null rays beginning just outside $\partial B$ do not reach the black hole horizon.    

The McVittie metric in the flat FRW universe is given by
\begin{equation}
    \begin{split}
        ds^{2}=-\left({\frac {1-{\frac {M}{2a(T)\mathcal{R}}}}{1+{\frac {M}{2a(T)\mathcal{R}}}}}\right)^{2}dT^{2}+\left(1+{\frac {M}{2a(T)\mathcal{R}}}\right)^{4}a^{2}(T)\left(d\mathcal{R}^{2}+\mathcal{R}^{2}d\Omega ^{2}\right).
    \end{split}
\end{equation}
It is easy to show that when taking
\begin{equation}
    \begin{split}
        x = a(T)\left(1+{\frac {M}{2a(T)\mathcal{R}}}\right)^2\mathcal{R},
    \end{split}
\end{equation}
the McVittie metric becomes
\begin{equation}\label{eq:mcvittie}
    \begin{split}
        ds^2 = -\left(1-\frac{R_s}{x}-H^2x^2\right)dT^2 -\frac{2Hx}{\sqrt{1-\frac{R_s}{x}}}dT\, dx + \frac{dx^2}{1-\frac{R_s}{x}} + x^2d\Omega^2.
    \end{split}
\end{equation}
In the static limit where the Hubble parameter $H=0$, this is exactly the Schwarzschild metric. The shell $\partial B$ is static for the exact Schwarzschild case as shown in \cite{mukhanov,mukhanov1}, but the $dT dx$ term in Eq.(\ref{eq:mcvittie}) makes the shell collapse. 

We will treat only McVittie black holes with Schwarzschild radii much smaller than the particle horizon of the $p = \rho$ universe. Thus, to first approximation we can treat the black hole as a static black hole in Minkowski space. We want to ask whether it is possible to embed a horizon volume of the asymptotically dS universe given by the flat FRW metric with scale factor $\sinh^{1/3} (3t/R_n)$, into the interior of the black hole, by matching along some time-like surface satisfying the Israel junction conditions with surface stress tensor obeying the dominant energy condition (DEC).  

To do this, we try to work in asymptotically static coordinates, where the AsdS metric is approximately
\begin{equation}\label{eq:dS}
    ds^2 = -  \left(1 - \frac{r^2}{R_n^2}\right) dt^2 + \frac{dr^2}{1 - \frac{r^2}{R_n^2}}  + r^2 d\Omega^2 . 
\end{equation} 
The exterior Schwarzschild metric is
\begin{equation}\label{eq:schwarz}
    ds^2 = 
    - \left(1- \frac{R_s}{x}\right)dT^2  + \frac{dx^2}{1- \frac{R_s}{x}}  + x^2 d\Omega^2. 
\end{equation} 
We want to match the induced metrics along a shell satisfying the Israel junction conditions with a boundary stress tensor obeying the dominant energy condition (DEC).  

The coordinates on both sides of the time-like shell become functions of the proper time $\tau$ on the shell.  Denoting the coordinate positions of $\partial B$ by $r(\tau)$ and $x(\tau)$, 
the condition that the matching spheres have the same area is
\begin{equation} 
    r(\tau) = x (\tau) . 
\end{equation} 

The energy density $\kappa$ on the boundary must compensate for the difference in extrinsic curvatures.  This gives 
\begin{equation} 
    \kappa = -\frac{1}{4\pi r (\tau)} \left(\sqrt{\dot{r}^2 + 1 - \frac{R_s}{r}} -\sqrt{\dot{r}^2 + 1 - \frac{r^2}{R_n^2}}\right), \\
\end{equation}  
where $\dot{r} = dr/d\tau$, see Appendix \ref{app:app1} for calculations. 
If the shell is space-like, the surface energy density picks a minus sign. In both cases, we can rewrite the above equation as 
\begin{equation} 
\begin{split}
    \dot{r}^2 - \frac{r^2}{R_n^2} -\left(\frac{ \frac{R_s}{r} - \frac{r^2}{R_n^2}}{8\pi \kappa r} + 2\pi \kappa r\right)^2 + 1   = 0 .
\end{split} 
\end{equation} 
The critical radius $r_c$ with a vanishing surface energy density is 
\begin{equation}
    \begin{split}
        r_c = \left(R_s R_n^2 \right)^{1/3}.
    \end{split}
\end{equation}
When $r>r_c$, $\kappa < 0$, the shell expands, and when $r<r_c$, $\kappa>0$, the shell collapses. The surface pressure/tension of the shell is given by 
\begin{equation} 
    p = \frac{1}{8\pi r} \left[ \frac{\dot{r}^2 + \ddot{r}r + 1 - \frac{3R_s}{2r}}{\sqrt{\dot{r}^2 + 1 - \frac{R_s}{r}}} - \frac{\dot{r}^2 + \ddot{r}r + 1 }{\sqrt{\dot{r}^2 + 1 - \frac{r^2}{R_n^2}}}\right] . 
\end{equation}  
It is straightforward but tedious to verify that $\kappa \geq |p|$. 

Following Frolov and Mukhanov, the behavior of the shell depends on whether its velocity is space-like or time-like. For time-like velocity, the dominant energy condition implies that the shell is collapsed by the shrinking Schwarzschild geometry. We interpret this as thermalization of the dS universe's q-bits with those of the larger entropy black hole. For space-like velocity, the shell crosses the Einstein-Rosen bridge and emerges on the other side into a white hole geometry. The Israel condition turns this into an asymptotically dS Big Bang universe, connected to the black hole by an ER bridge. By the usual ER=EPR logic \cite{ER=EPR}, this is interpreted as a description of the entanglement of the low entropy dS subsystem with the larger black hole system which has subsumed it.

From the point of view of a detector sitting at the origin of dS space,  the beginning of the collapse of the shell is detected as soon as light can travel from the boundary of the shell to the detector, which is a detector's proper time of order $R_n {\rm ln} (R_n /\delta_n) $, where $\delta_n$ is the distance of the shell's initial position from the dS horizon.  The logarithm represents the effect of the redshift of near horizon signals as seen by the detector. For a time between this and $R_s$ the effects of the collapse become more pronounced and at a time of order $R_s$ the detector ``hits the black hole singularity".  Our interpretation of this is in terms of entropy.  For $R_s > R$ the asymptotically dS space is a subsystem of the larger black hole system, which is kept causally isolated until detector times of order $R_n {\rm ln} (R_n /\delta_n)$ , but then begins to mix and equilibrate with the larger system.  This is seen geometrically by the fact that as time goes on the detector at the origin can explore causal diamonds of smaller and smaller area.  Degrees of freedom previously accessible to it have become absorbed in the large black hole horizon.  The first signs of this become evident to a bulk detector at roughly the scale of the lifetime of the most robust possible bulk detector.  The details of the collapse and time-scale for final equilibration depend on $R_s$, which is always invisible to any bulk detector if $R_s \gg R_n$.  Eventually, the McVittie black hole will decay into the infinity entropy $p = \rho$ background, but this is even more remote from bulk observations. 

We can now understand that the $\epsilon < 0$ branch of solutions \cite{mukhanov,mukhanov1}, or at least its stationary asymptote, as describing the entanglement of the dS subsystem with the black hole, in the equilibrium state of the latter \cite{ER=EPR}.  This is further evidence for the interpretation of the collapse of localized systems to black hole singularities as equilibration of their degrees of freedom with the larger black hole horizon.

The $p = \rho$ cosmology has an infinite asymptotically null boundary,  conformal to the future half  of the boundary of Minkowski space.  Two of us have proposed a quantum theory for it in \cite{tbwflm}, but we don't want to assume that that model is correct in this paper\footnote{In fact, that model applies to the cosmologies with general values of $R_n$ and contradicts the claims the asymptotically dS universes have only a single state.  We are attempting to make a model independent conclusion in the current paper.}.  While the nature of the quantum theory describing such a universe is not a settled issue, it seems clear from any point of view that it must posses an infinite dimensional Hilbert space.  The McVittie black holes should be viewed as finite entropy subsystems in that Hilbert space, and for $R_s > R$ the asymptotically dS universes are finite entropy subsystems of them.  In this context it is clear that the ``closed" asymptotically dS universe does not have a one dimensional Hilbert space.

Finally we should note that if we have multiple McVittie black holes in a $p = \rho$ universe, then, depending on initial conditions, pairs of them might collide on time scales shorter than those considered above.  There is no way to tell, from observations inside one of the interior asymptotically dS universes, what the initial positions and velocities in  the multiple black hole multiverse were.  We have not worked out the consequences of these collisions for the interiors, but they must be catastrophic.  Causality assures us that the catastrophe will not occur on a time scale shorter than $R_n$, but anything longer than that is pure speculation.

\section{Biothropic Considerations}

String theory has provided us with an enormous amount of evidence that there are many mathematically consistent models of quantum gravity, characterized by a wide variety of parameters, among which is a discretely tuneable parameter that fixes the cosmological constant.  Well established models all have non-positive values of the the c.c..  The conjectures of \cite{tbwfds,tbwfds1,tbwfds2}, which have been subsumed under the generalized \cite{BZ} Carlip-Solodukhin \cite{carlip,solo} conjecture, imply that the same is true for positive c.c..  The model we have sketched above allows us to incorporate all possible models with positive c.c. into a single ``multiverse" model.  We would like to emphasize that unlike models based on conjectures about ``Eternal Inflation", our model is, at any finite time, a finite unitary quantum mechanics, that is built to incorporate causality and at least a coarse-grained notion of relativity.  

A number of things about our model are noteworthy, when compared to previous attempts to explain the value of the c.c. in terms of preservation of conditions under which intelligent life could develop.  First of all, all dS universes have finite life-times, with those having c.c. close to the Planck scale lasting only for a Planck time.  Secondly, in a previous paper \cite{satbcmb}, we have presented a comprehensive theory of the CMB and dark matter in an asymptotically dS universe.  In order to create a radiation dominated universe at all, we need an initial universe containing many horizon volumes filled with black holes of Schwarzschild radius $\sim 10^6 L_P$.  The precise number was fixed to fit the observed fluctuations in the CMB, but if the black holes had been much smaller there would have been much larger density fluctuations and the black hole dominated era might not have ended at all.  We also needed a random distribution of much larger primordial black holes to serve as dark matter.

Weinberg argued \cite{weinbergcc} that in order to have life, all we need is one galaxy.  In the very specific scenario where black holes of various sizes give rise both to the CMB and to the dark matter, then the story of galaxy formation may be more complex and may give us more of a constraint than just an upper bound on the c.c..

All string theory models in asymptotically flat space are exactly supersymmetric.  Once one accepts that the c.c. is a tunable parameter, it is natural to expect that there might be sequences of dS models, which converge to supersymmetric models in Minkowski space.  This also leads one to conjecture connections between supersymmetry (SUSY) breaking and the value of the c.c. \cite{tbsusy}.  Whatever the validity of those particular conjectures, SUSY breaking must go to zero with the c.c..  Exact supersymmetry is a disaster for both nuclear physics and chemistry.  The periodic table and the shell model disappear.  Stars behave very differently when fermi pressure disappears and the lowest states become Bose condensates.  So, if there is any sort of power law connection between the c.c. and the strength of SUSY breaking, there is a lower bound on the c.c. that comes from demanding the existence of {\it life of our type}.  We italicize that phrase, because it signals a problem with almost all kinds of arguments based on the existence of ``intelligent life".   We don't understand enough of the chemistry and biology of our own universe to know what kind of intelligent life is possible under what conditions.  For a universe with different low energy gauge interactions we probably couldn't even figure out the rudiments of ``nuclear physics" and ``chemistry".   This is why only purely qualitative gravitational arguments, like Weinberg's ``one galaxy" requirement are convincing environmental selection principles.  We are not even completely convinced that a ``supersymmetry precludes the existence of life" argument is valid.  It certainly rules out the kind of chemistry and nuclear physics that's required for life of our type.  

By it is very nature, a multiverse explanation of some fact via environmental selection is stretching the boundaries of real science because the other universes in the multiverse can't be probed by experiment.  Thus, we prefer to have the minimal number of parameters and initial conditions in our models ``explained" by biothropic principles\footnote{Andy Albrecht has formalized this in the epigram, "The physicist who has the fewest anthropically determined parameters in their theory wins".}.  Instead, it is probably better to just outline what conditions in the known universe require one to do something that is not the most obvious or natural thing from the point of view of the mathematics of the model.   

\section{Conclusions}

Speculations that closed universes are properly modeled by a single dimensional quantum Hilbert space, or that a universe that appears to be approaching an asymptotically dS state in the future, will experience recurrences, cannot be justified by any observation made within that universe.  There are models where such a universe is a finite dimensional subsystem of a much larger system, which equilibrates with the larger system on a time scale longer than the lifetime of any robust detector of quantum information in the dS universe, but much shorter than a recurrence time.  These models provide a framework for environmental selection of physical constants like the c.c., as well as cosmological initial conditions.  

We would like to emphasize that our previous paper \cite{galaxyseeds} attempted to explain the coarse grained properties of our own universe based on a mini version of the multiverse model described here.  There, we again utilized the Israel junction condition to embed asymptotically dS maximal entropy universes of various sizes into black holes. However, in order to have ``ordinary" radiation and matter dominated eras, we assumed that most of the black holes were very tiny, so that they decayed into particles very early in the history of the universe.  A distribution of larger black holes then accounted for the dark matter that led to galaxy formation.  The key fact about quantum gravity that we were exploiting is the lesson taught to us by the Schwarzschild de Sitter entropy formula and old arguments about the limitations of quantum field theory.  All localized excitations in an asymptotically dS universe are low entropy states, and the maximal entropy for given energy is in black holes.  In a given causal diamond, only a sub-leading power of the entropy can be described by quantum field theory.  The multiverse model described in this paper is just a grandiose version of that single universe model, without the constraints of accounting for observational data.  Its purpose was to show that observational data cannot be extrapolated to make unwarranted philosophical conclusions about the universe we live in, and to provide a framework in which some ``fundamental constants" could be determined by environmental selection within a controllable mathematical framework.  We have described our model, for the most part, in terms of solutions to Einstein's equations, but there are finite quantum mechanical systems whose hydrodynamics mimics these equations closely.  

Finally, we would like to end with a comment about the use of the phrase ``closed universe" to describe asymptotically dS cosmologies.  This terminology is motivated by the global coordinate system on dS space, which has closed space-like slices.  Experimental physics is however always tied to what can be measured by a single detector (whatever that word might mean) traveling on a time-like or null trajectory in space-time.  AsdS spaces always have at least two maximal causal diamonds with only partial asymptotic overlap.  Since the pioneering work of Israel \cite{IsraelTFD} we have learned to interpret such space-times in terms of distinct entangled quantum mechanical systems.  This point of view has been rigorously validated in the AdS/CFT correspondence \cite{MaldaTFD}.  Thus, any attempt to interpret an AsdS universe as a single quantum system is contradicting this piece of conventional wisdom.  Once one accepts that the correct quantum mechanical point of view about these systems is to concentrate on the quantum mechanics measurable by a detector in one of the entangled systems, then the AsdS universe is not closed.  Of course, this is the conclusion come to, in a rather roundabout way, by at least some of the authors of papers claiming that a closed universe has a single state.  One then has to deal with the intrinsic quantum gravitational limitations on any detector, which we have discussed in the introduction.  

\appendix
\section{Calculations for surface energy density and surface pressure}\label{app:app1}
In this appendix, we find the surface energy density $\kappa$ and the surface pressure $p$ of the shell following the Israel junction condition. Consider a general metric of the following form
\begin{equation}
    \begin{split}
        ds^2 &= -f(r)dt^2 + \frac{dr^2}{f(r)} + r^2d\Omega^2, \\
        g_{\mu\nu} &= \text{diag}\left(-f(r),1/f(r), r^2,r^2\sin^2\theta\right).
    \end{split}
\end{equation}
The induced metric on the thin time-like shell with radius $r(\tau)$ is given by
\begin{equation}
    \begin{split}
        ds^2 = -d\tau^2 +r^2(\tau)\left(d\theta^2+\sin^2\theta \, d\phi^2\right),
    \end{split}
\end{equation}
where $\tau$ is the proper time on the shell, and we use dot to denote the derivative with respect to $\tau$. The 4-velocity $u^{\mu} = \left(\dot{t},\dot{r},0,0\right)$ satisfies the normalization condition $u_{\mu}u^{\mu}=-1$, this gives
\begin{equation}
    \dot{t} = \frac{\sqrt{\dot{r}^2+f(r)}}{f(r)}.
\end{equation}
The nonzero components of the extrinsic curvature tensor $K_{ij}$ are
\begin{equation}
    \begin{split}
        {K^{\tau}}_{\tau} &= -\frac{\ddot{r}+f'
        (r)/2}{\sqrt{\dot{r}^2+f(r)}}, \hspace{20mm}
        {K^{\theta}}_{\theta} = {K^{\phi}}_{\phi} = \frac{\sqrt{\dot{r}^2+f(r)}}{r},
    \end{split}
\end{equation}
where $n^{\mu}$ is the unit normal satisfying $n_{\mu}u^{\mu}=0$. The surface energy density $\kappa$ and surface pressure $p$ are given by the difference between the extrinsic curvatures on the outside and inside of the shell, which we denote using indices $\pm$. The metrics of dS exterior and Schwarzschild interior are given in Eq.(\ref{eq:dS}) and (\ref{eq:schwarz}), bringing $f(r)_{\pm}$ into the above expressions, we find 
\begin{equation}
    \begin{split}
        \kappa &= -\frac{1}{4\pi}\left[\left({K^{\theta}}_{\theta}\right)_+-\left({K^{\theta}}_{\theta}\right)_-\right] \\
        &= -\frac{1}{4\pi r (\tau)} \left(\sqrt{\dot{r}^2 + 1 - \frac{R_s}{r}}  - \sqrt{\dot{r}^2 + 1 - \frac{r^2}{R_n^2}}\right).
    \end{split}
\end{equation}
Similarly, surface pressure $p$ is given by
\begin{equation}
    \begin{split}
        p &= \frac{1}{8\pi}\left[({K^{\theta}}_{\theta})_+-({K^{\theta}}_{\theta})_-+\left({K^{\tau}}_{\tau}\right)_+-\left({K^{\tau}}_{\tau}\right)_-\right] \\
        &= \frac{1}{8\pi r} \left[ \frac{\dot{r}^2 + \ddot{r}r + 1 - 3R_s / 2r}{\sqrt{\dot{r}^2 + 1 - R_s/r}} - \frac{\dot{r}^2 + \ddot{r}r + 1 }{\sqrt{\dot{r}^2 + 1 - r^2/R_n^2}}\right] .
    \end{split}
\end{equation}

\acknowledgments

The work of S. A is supported in part by the DOE under grant DE-SC0010008. We thank H. Green for pointing out an error in the calculations in an earlier version of this manuscript.





\end{document}